# Simultaneous Calibration and Navigation (SCAN) of Multiple Ultrasonic Local Positioning Systems

David Gualda*, Jesús Ureña, Juan C. García, Enrique García, José Alcalá

*Department of Electronics (University of Alcalá), Alcalá de Henares, Madrid, Spain.*
*david.gualda@depeca.uah.es*

**Abstract**

This paper proposes a Simultaneous Calibration and Navigation (SCAN) algorithm of a multiple Ultrasonic Local Positioning Systems (ULPSs) that cover an extensive indoor area. The idea is the development of the same concept than SLAM (Simultaneous Localization and Mapping), in which a Mobile Robot (MR) estimates the map while it is navigating. In our approach, the MR calibrates the beacons of several ULPSs while it is moving inside the localization area. The concept of calibration is the estimation of the position of the beacons referenced to a known map. The scenario is composed of some calibrated ULPSs that we denote as Globally Referenced Ultrasonic Local Positioning Systems (GRULPSs) that are located in strategic points like entrances covering the start and the end of a possible trajectory in the environment. Additionally, there are several non-calibrated ULPSs named Locally Referenced Ultrasonic Local Positioning Systems (LRULPSs) that are placed around the localization area. The proposal uses a MR with odometer for calibrating the beacons of the LRULPSs while it is navigating on their coverage area and go from one GRULPS to another. The algorithm is based on multiple filters running in parallel (one filter for each LRULPS and another one for the GRULPSs) that estimate the global and local trajectories of the MR (one trajectory for each local reference system of the LRULPSs) fusing the information related to the Ultrasound Signals (US) and the odometer of the MR. The position of the beacons of the LRULPSs are obtained by a transformation vector for each LRULPS that converts the local coordinates to the global reference system. This transformation vector is calculated using several points of a local trajectory and the corresponding points of the global one. The method is independent of the type of filter, provided that it works properly with non-linear systems and possibly non-Gaussian noise. Extended Kalman Filter (EKF), Unscented Kalman Filter (UKF) and H-∞ Filter have been tested, in simulations and real experiments, in order to compare their performance in this case.

## 1. Introduction

Since the GPS appeared and solved the outdoor localization, the challenge is the development of a similar system for indoor environments in terms of accuracy and availability.

The most extended technology to develop an indoor location system is Radiofrequency (RF) due to the low impact in the building infrastructure using the Access Points (APs) of the WiFi service [1][2][3]. The problem is that the accuracy of these kind of systems is not as ideal as the GPS due to the propagation characteristics of the RF signal in indoor environments. The majority of works that use this technology report mean errors of few meters with a high variance [4][5]. Nevertheless, some works present an accuracy of decimeters and centimeters deploying a high number of nodes and short-range communications [6], measuring the Time of Flight (ToF) with a nano-second resolution [7], or using multiple WiFi antennas [8].

Another technology less extended than RF is Ultrasound (US), whose advantage is its accuracy (centimeters in the most of works) due to the low propagation speed. Nonetheless, the disadvantage is the need of installing an external infrastructure and the reduced coverage area of the US signals (few meters). Each US infrastructure is composed of several beacons that form an Ultrasonic Local Positioning Systems (ULPS) [9][10]. Since its development, some upgrades appeared in order to improve the performance, such as the inclusion of codes in the emission of the beacons to avoid crosstalk and allow them to emit simultaneously [11].

For covering indoor extensive areas, a solution is the deployment of several ULPSs such as in [12]. Nevertheless, one of the problems is the calibration of the beacons of each ULPS, that is, the estimation of the position of each beacon related to the global map. Usually, the process for calibrating an Indoor Localization System (ILS) composed of a single LPS is manual and it is based on the use of a plumb to determinate the beacons' projection on the floor and solve the 2D beacons' position using several distances from the projected points to some known reference positions of the environment like corners or different points on a wall. The height of the beacon is usually obtained by a laser meter. Other method that requires less effort than the first one is the use of the inverse positioning. The process is based on capturing several distance measures in known points on the floor (i.e. using a grid) and then, obtaining the position of the beacons using a localization algorithm [13][14]. In this case, it is necessary to know accurately the positions of a high number of measurement points. Normally, the selection of the floor points is arbitrary, but in [15] an algorithm determines the optimal points for beacons' calibration, minimizing the error. Recently, other method completely autonomous is based on the use of a MR with odometry for calibrating one LPS [16].

MRs are used in several researches for developing a map of indoor environments while they are navigating. This process is called Simultaneous Localization and Mapping (SLAM) [17][18][19], in which the robots are equipped with several distance meters to obtain measurements from the robot to the objects (walls, pillars, …) of the environment. Using these distances it is possible to make an estimation of the the the map while the robot is navigating and also computes its own trajectory. In this paper we propose the use of the SLAM concept to the autonomatic calibration of US beacons (that is, obtaining the coordinates of the beacons' positions) applying the method to multiple ULPSs instead of only to one [16]. We use the distances or difference of distances between the MR and the beacons to estimate the position of them. These distances are measured while the robot is navigating around the area of interest in which there are some non-calibrated ULPSs. Therefore, this proposal allows us calibrating multiple ULPSs automatically with a MR, avoiding the need of tedious manual methods. We denote the proposal as Simultaneous Calibration and Navigation (SCAN).

Regarding localization algorithms, for static positions or for the first position of the MR one of the most extended in literature is the Gauss-Newton one [20] for solving the non-linear equation system based on the distances between the receiver located on the MR and each emitter (beacon). These distances can be measured with ultrasound if there is a synchronization between the emitters and the receiver (spherical case).

If there is no synchronization, then, the difference of distances, using one of the beacons as reference, are computed (hyperbolic case) [21].

In order to update the position of the MR fusing the available information (ultrasound signals and odometry in our case) it is very common the use of a filter [22]. The most representative for dealing with non-linear problems are Extended Kalman Filter (EKF) [23], Unscented Kalman Filter (UKF) [24], H-∞ filter [25] and Particle Filter (PF) [26]. The last one is more appropriate when the measures present a high variance and an unknown noise distribution such as localization systems based on radiofrequency (RF) or combination of different technologies [27]. For fusing information of measures based on ultrasound signals the most recommendable are EKF, UKF, H-∞ and other variations of them since the measures of these signals present a lower variance than the ones from RF and their noise distribution can be approximate to a zero-mean Gaussian distribution in some cases (not multipath).

As for the organization of the paper, section 2 describes the architecture of the system and the objectives to be achieved; Section 3 shows the development of the proposed solutions describing the equations of the filters; Section 4 shows the simulated results; Section 5 shows the real experiments to validate the proposal; and, finally, Section 6 deals with some conclusions.

## 2. Statement of the Problem

As described in the introduction, an ULPS allows locating a target with a high accuracy in a reduced zone. If an application requires locating a target in an extensive area using the ultrasound technology, the solution is the deployment of a set of ULPSs covering the full area for localization and navigation tasks. In our proposal we define two kind of ULPSs (Fig. 1). Globally-Referenced Ultrasonic Local Positioning System (GRULPS), in which the positions of its beacons $\mathbf{B}_{G,m}$ are known and referenced to the global reference system of the map, and they have been, for instance, manually calibrated. We suppose the errors of this method can be negligible, as they could be around millimeters using a laser plumb, that is below the error of distances measurement with ultrasounds. These GRULPSs can be located in strategic zones like the entrances of the environment, and Locally-Referenced Ultrasonic Local Positioning System (LRULPS), whose beacons' position $\mathbf{B}'_{L,n}$ are unknown with respect to the global reference system. On the other hand, in a LRULPS the arrangement of the beacons inside its structure is known and then we can reference their position regarding a local reference system (this locally-referenced positions will be included in the matrix $\mathbf{B}_{L,n}$). Several of these uncalibrated LRULPSs will be located covering areas between GRULPSs and the number and position of them can be changed along the time. Note that the sub-indexes G and L mean "global" and "local" respectively, while $m = 1, 2, \ldots, M$ and $n = 1, 2, \ldots, N$ denote the number of a particular GRULPS or LRULPS, respectively.

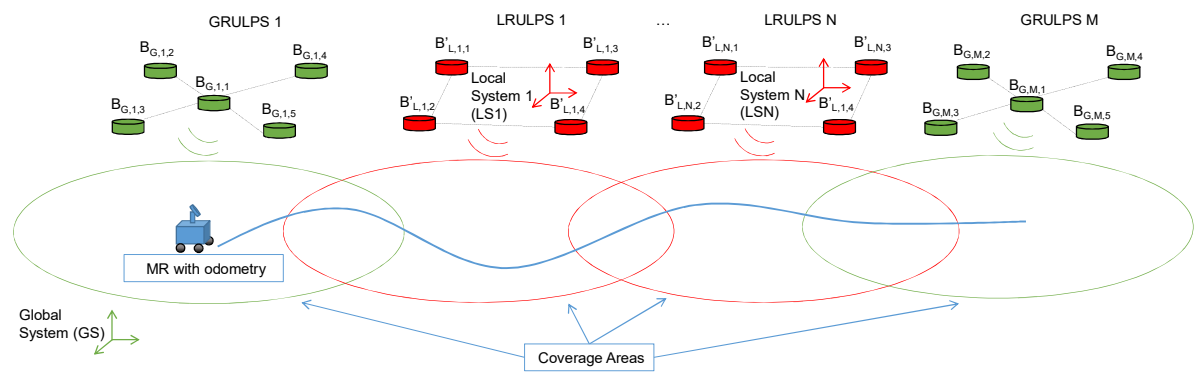

**Fig. 1.** Particular case of GRULPS and LRULPS structures, with five and four beacons, respectively.

Each ULPS has to be composed of at least three beacons in the spherical positioning case (working with distances between the emitters and the receiver) or four beacons in the hyperbolic one (working with difference of distances).

The contribution of this paper is the development of an algorithm that allows calibrating all LRULPS beacons of the proposed structure with a Mobile Robot (MR) at the same time that it navigates and its trajectory is being obtained.

Fig. 2. shows the simulated scenario, where all beacons are placed on the ceiling, at a height of 3.5 m. The representation includes the projection of the beacons on the floor, for both, GRULPSs (M=2) and LRULPSs (N=7), and their coverage areas on the ground. In this case, GRULPSs are composed of five beacons and LRULPSs of four beacons, therefore GRULPSs have more redundancy. Because of the performance achieved with the current ULPS used in real tests [28], we assume that for all ULPSs, the coverage area is a circle on the ground centered at the projection of the center of the ULPS on the floor with a radius of five meters. The structure of each ULPS is squared with a diagonal of 1m. Some kind of overlapping between the ULPSs is recommended in order to reduce the positioning error of the MR and the LRULPS beacons' position estimation. Note that the highest error in the estimation of the beacons' positions corresponds to the furthest LRULPS with respect to a GRLPS (e.g. LRLPS #4 in Fig. 2). If the MR moves outside the ULPS coverage area (either a GR or LR ULPS) it will only estimate its position using the odometry information, so the error will increase accumulatively.

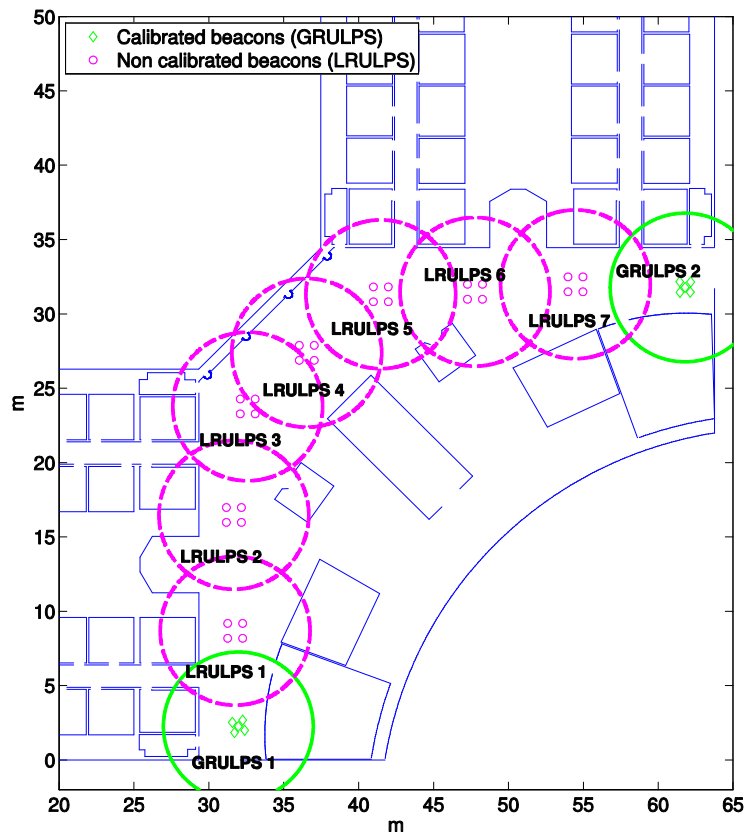

**Fig. 2.** Simulated workspace based on the real plan of a building.

### 3. Simultaneous Calibration and Navigation (SCAN)

#### 3.1. Introduction

The proposal estimates the 2D location and orientation (pose) of a MR with odometry inside a global coordinate system. The estimated pose (position $(\hat{x}_{G,k}, \hat{y}_{G,k})$, and orientation $(\hat{\theta}_{G,k})$) for the current instant $k$ is described by the global state vector as:

$$\hat{\mathbf{X}}_{G,k} = \begin{bmatrix} \hat{x}_{G,k} & \hat{y}_{G,k} & \hat{\theta}_{G,k} \end{bmatrix}^T \tag{1}$$

When the MR is located inside the coverage area of a LRULPS, another local state vector is defined, in addition to the global one, as:

$$\hat{\mathbf{X}}_{L,n,k} = \begin{bmatrix} \hat{x}_{L,k} & \hat{y}_{L,k} & \hat{\theta}_{L,k} \end{bmatrix}^T \tag{2}$$

$n = 1, 2, \ldots, N$ is the specific LRULPS.

In order to have always a globally referenced position and orientation of the MR, let us suppose that the first iterations of the MR will be inside the coverage area of a GRULPS. This is a convenient condition and it is also quite easy to accomplish, since GRULPSs have been specified to be installed on the entrance or exit points of the main section areas of interests.

While the position of the MR is being estimated, the proposal also obtains the LRULPS beacons' coordinates referenced to the global map using a transformation vector between reference systems. When a LRULPS is being calibrated, the method uses that LRULPS as a new GRULPS and estimates the MR position according to its reference system. The 2D position of each beacon is represented in this local reference system as:

$$\hat{\mathbf{B}}'_{L,n,i} = \begin{bmatrix} \hat{x}'_{B,n,i} & \hat{y}'_{B,n,i} \end{bmatrix}^T \tag{3}$$

The sub-index $i = 1,2,\ldots,I$ represents a beacon of the $n\text{-}th$ LRULPS and the sub-index B means "beacon".

Fig. 3 and Fig. 4 show the process of estimating the MR position inside the GRULPSs and LRULPSs coverage areas and the following subsections explain the main blocks of the proposed method.

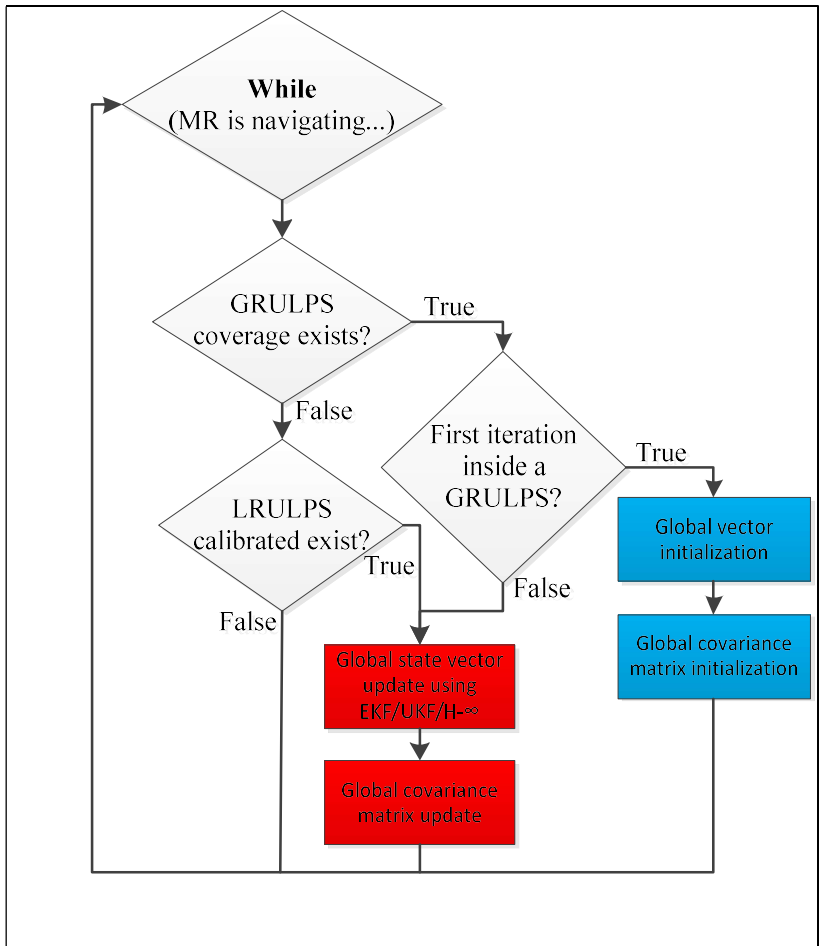

**Fig. 3.** Flowchart of the estimation of the MR position inside GRULPSs and already calibrated LRULPSs.

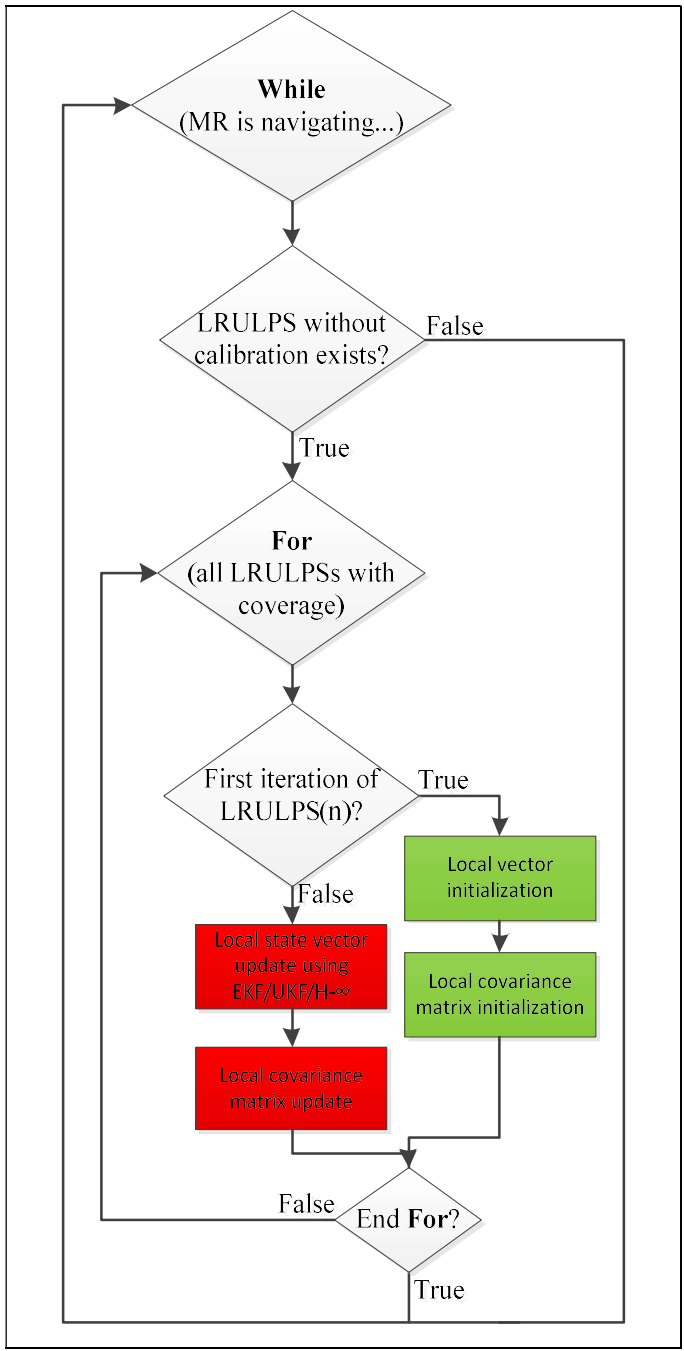

**Fig. 4.** Flowchart of the estimation of the MR position inside LRULPSs without calibration.

### 3.2. Global vector initialization (blue blocks of Fig. 3)

When the MR goes into a GRULPS area for the first time (e.g. in the first iteration), the global state vector $\hat{\mathbf{X}}_{G,0}$ is obtained using a Gauss-Newton (GN) minimization algorithm. It solves the position of the MR using the measures obtained with the ULPS, that is, the distances between the MR and the beacons $d_{\mathbf{P}_{G,0},\mathbf{B}_{G,m,i}}$ or the difference of distances $\Delta d_{\mathbf{P}_{G,0},\mathbf{B}_{G,m,i}}$ respect to a reference beacon, depending on the case (spherical or hyperbolic). The estimated position of the MR at the first iteration using the GN algorithm is:

$$\hat{\mathbf{P}}_{G,0} = [\hat{x}_{G,0} \quad \hat{y}_{G,0}]^T \tag{4}$$

The estimated distances or differences of distances are given by the expressions (5) and (6), respectively:

$$\hat{d}_{\hat{\mathbf{P}}_{G,0},\mathbf{B}_{G,m,i}} = \sqrt{(\hat{x}_{G,0} - \hat{x}_{B,m,i})^2 + (\hat{y}_{G,0} - \hat{y}_{B,m,i})^2 + (z_{MR} - z_{B,m,i})^2} \tag{5}$$

$$\Delta\hat{d}_{\hat{\mathbf{P}}_{G,0},\mathbf{B}_{G,m,i}} = \begin{aligned}&\sqrt{(\hat{x}_{G,0} - \hat{x}_{B,m,i})^2 + (\hat{y}_{G,0} - \hat{y}_{B,m,i})^2 + (z_{MR} - z_{B,m,i})^2}\\&-\sqrt{(\hat{x}_{G,0} - \hat{x}_{B,m,1})^2 + (\hat{y}_{G,0} - \hat{y}_{B,m,1})^2 + (z_{MR} - z_{B,m,1})^2}\end{aligned} \tag{6}$$

All the variables have been previously defined. The sub-index $m = 1, 2, \dots, M$ represents the specific GRULPS. Note that in the hyperbolic case, $i = 2,3, \dots, I$ because the first beacon is the reference one. The height of the MR is a known constant value, $z_{MR}$.

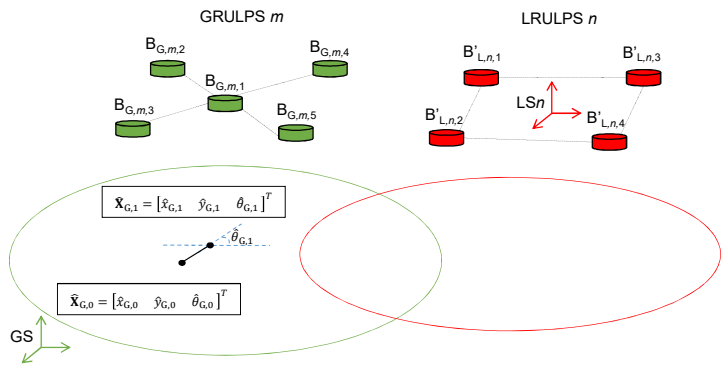

**Fig. 5.** Diagram of the first and second iterations of the MR in the global reference system.

The following step in order to complete the initialization of the global state vector is the computation of the initial orientation of the MR. When the MR moves, a second different MR position is obtained, using the GN algorithm, from a new set of ULPS measurements. With two MR consecutive positions we can calculate the MR orientation. Fig. 5 shows a diagram of the first and second iterations of the MR in the global reference system.

### 3.3. Local vector initialization (green blocks of Fig. 4)

When the MR moves inside a LRULPS coverage area, the first iteration of the $n^{th}$ local state vector $\widehat{\mathbf{X}}_{L,n,0}$ is obtained using the same procedure as for the GRULPS case. The estimated position of the MR, regarding the local reference system, at the first iteration of the LRULPS $n$ is defined as:

$$\widehat{\mathbf{P}}_{L,n,0} = [\hat{x}'_{L,n,0} \quad \hat{y}'_{L,n,0}]^T \tag{7}$$

In this case, the estimated distances or differences of distances are given by the expressions (8) and (9), respectively.

$$\hat{d}_{\widehat{\mathbf{P}}_{L,n,0},\mathbf{B}_{L,n,i}} = \sqrt{(\hat{x}_{L,n,0} - \hat{x}_{B,n,i})^2 + (\hat{y}_{L,n,0} - \hat{y}_{B,n,i})^2 + (z_{MR} - z_{B,n,i})^2} \tag{8}$$

$$\Delta\hat{d}_{\widehat{\mathbf{P}}_{L,n,0},\mathbf{B}_{L,n,i}} = \begin{array}{l} \sqrt{(\hat{x}_{L,n,0} - \hat{x}_{B,n,i})^2 + (\hat{y}_{L,n,0} - \hat{y}_{B,n,i})^2 + (z_{MR} - z_{B,n,i})^2} \\ -\sqrt{(\hat{x}_{L,n,0} - \hat{x}_{B,n,1})^2 + (\hat{y}_{L,n,0} - \hat{y}_{B,n,1})^2 + (z_{MR} - z_{B,n,1})^2} \end{array} \tag{9}$$

The sub-index $n = 1, 2, \ldots, N$ represents the specific LRULPS. Fig. 6 shows the diagram of the first iteration of the MR in the $n^{th}$ local reference system.

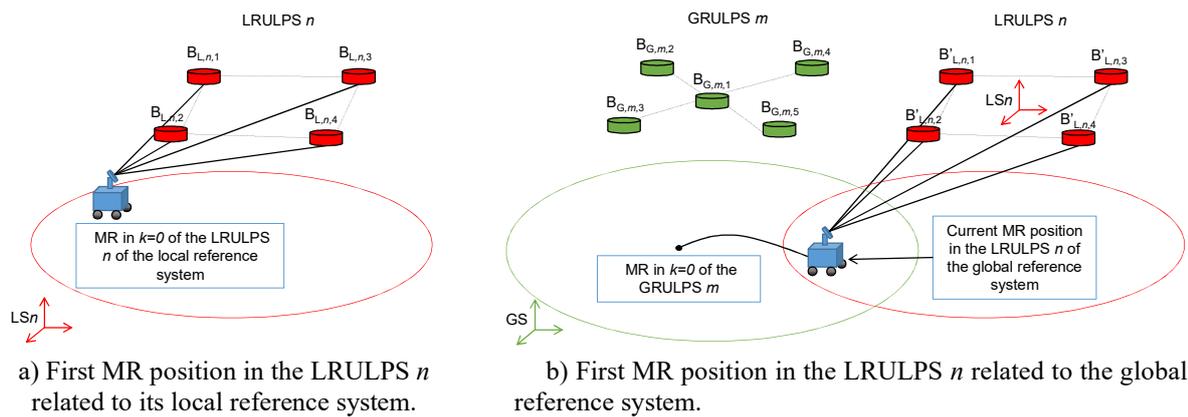

a) First MR position in the LRULPS $n$ related to its local reference system.

b) First MR position in the LRULPS $n$ related to the global reference system.

**Fig. 6.** Diagram of the first iteration of the MR in the $n^{th}$ local reference system.

The initial orientation of the LRULPS $n$ is also obtained using two consecutive MR position points. Fig. 7 shows the diagram.

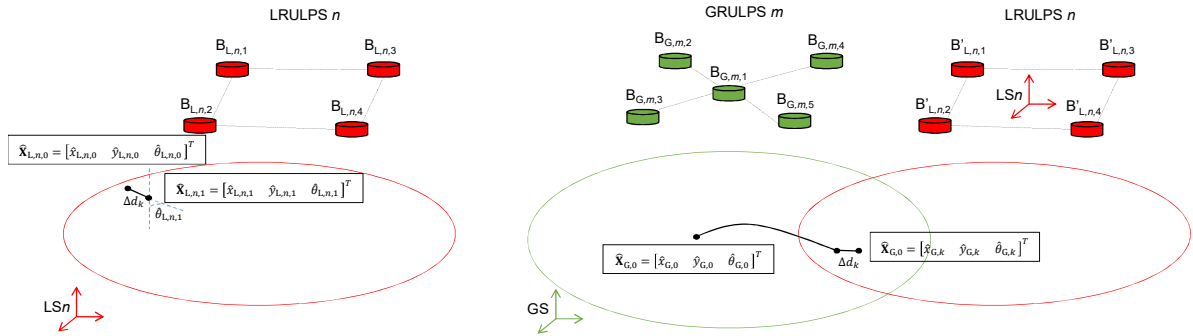

**Fig. 7.** Initialization of the local orientation using the first two points of the trajectory.

### 3.4. Global and local state vectors update using a filter (red blocks of Figs. 3 and 4)

For updating the global and local state vectors of the MR during the iterations we use a filter. This filter can be of any kind that estimates a state vector and its covariance matrix of a non-linear discrete process using a set of observations (US measurements in our case). Note that the sub-index T represents the type of LPS; that is, $T = \{G, L\}$ denotes that the equations are valid for the global (G) and local (L) cases, and the sub-index o represents de number or order of LPS, with $o = \{m, n\}$ that is, the equation is applied to the $m^{th}$ GRLPS or the $n^{th}$ LRULPS, depending on the case (global or local).

The equation process particularized to our problem is:

$$\mathbf{X}_{T,k} = f\big(\mathbf{X}_{T-1,k}, \Delta d_{odo,k}, \Delta \theta_{odo,k}\big) + \mathbf{W}_T \tag{10}$$

The values $\Delta d_{odo,k}$ and $\Delta \theta_{odo,k}$ are the increments of distance and angle between the iteration $k$ and $(k-1)$ provided by the MR odometer and $\mathbf{W}_T$ is the process noise. We assume that it is zero-mean Gaussian with a covariance matrix denoted as $\mathbf{Q}_T$.

The measurement equation can be expressed as:

$$\mathbf{Z}_{T,o,k} = h\big(\mathbf{X}_{T,k}\big) + \mathbf{V}_{T,o,k} \tag{11}$$

The vector $\mathbf{Z}_{T,o,k}$ contains the US observations converted to distances or difference of distances, depending on the case (spherical or hyperbolic) of the $m^{th}$ GRULPS ($m$=o) or the $n^{th}$ LRULPS ($n$=o) at instant $k$, and $\mathbf{V}_{T,o,k}$ is the measurement noise that is assumed zero-mean Gaussian with a covariance $\mathbf{R}_{T,o}$. Note that the process and the measurement noise must be independent and uncorrelated.

For updating the rest of iterations of the global and local state vectors we use a filter that fuses the information of the US signals and the relative information represented by the odometry. In this case we have implemented three representative filters: the Extended Kalman Filter (EKF), the Unscented Kalman Filter (UKF) and the H-Infinity Filter (H-∞). The EKF adapted to our problem was introduced in [29], so the following subsections focus on the equations related to the UKF and the H-∞ filters.

*3.4.1. Unscented Kalman Filter (UKF)*

The first step is the calculation of the $(2L + 1)$ sigma points around the previous estimation according to the following expressions:

$$\chi_{T,(n),k,0} = \hat{\mathbf{X}}_{T,(n),k-1} \tag{12}$$

$$\chi_{T,(n),k,j} = \hat{\mathbf{X}}_{T,(n),k-1} + \left( \sqrt{(L + \lambda) \cdot \mathbf{P}^a_{T,(n),k}} \right)_j , j = 1, \dots, L \tag{13}$$

$$\chi_{T,(n),k,j} = \hat{\mathbf{X}}_{T,(n),k-1} - \left( \sqrt{(L + \lambda) \cdot \mathbf{P}^a_{T,(n),k}} \right)_{j-L} , j = L + 1, \dots, 2L \tag{14}$$

Note that the sub-index $(n)$ represents the $n^{th}$ reference system of the LRULPS when $T = L$. For the global case $(T = G)$ the parameter $(n)$ is not necessary, since the reference system is the same for all GRULPSs.    $L$ is the number of states (three in this case) and $\lambda$ is a scale factor which is equal to:

$$\lambda = \alpha^2 \cdot (L + \kappa) - L \tag{15}$$

The parameter $\alpha$ is a constant that determines the dispersion of the sigma points nearby the previous estimation of the state vector and, normally, is set to a low positive constant lower than 1. The other constant $\kappa$, is a secondary scale factor whose value is in the range of 0 to $(3 - L)$. The term $\left( \sqrt{(L + \lambda) \cdot \mathbf{P}^a_{T,(n),k}} \right)_j$ is the $j^{th}$ column of the square root matrix of $(L + \lambda) \cdot \mathbf{P}^a_{T,(n),k}$. We use a Cholesky decomposition in order to obtain the expression of the square root. $\mathbf{P}^a_{T,(n),k}$ is the augmented covariance matrix for the estimation and it is composed of the previous covariance matrices. The process noise and the measurement noise covariance matrices are as the following expression indicates:

$$\mathbf{P}^a_{T,(n),k} = \begin{bmatrix} \mathbf{P}_{T,(n),k-1} & 0 & 0 \\ 0 & \mathbf{Q}_T & 0 \\ 0 & 0 & \mathbf{R}_T \end{bmatrix} \tag{16}$$

Where the process noise $\mathbf{Q}_T$ is modeled by the following diagonal matrix:

$$\mathbf{Q}_T = \begin{bmatrix} \sigma^2_{w_{T,x}} & 0 & 0 \\ 0 & \sigma^2_{w_{T,y}} & 0 \\ 0 & 0 & \sigma^2_{w_{T,\theta}} \end{bmatrix} \tag{17}$$

The values $\sigma^2_{w_{T,x}}$, $\sigma^2_{w_{T,y}}$ and $\sigma^2_{w_{T,\theta}}$ are the variances related to each value of the global state vector. Regarding the measurement noise $\mathbf{R}_T$, in the spherical case its covariance matrix is defined in (18), whose diagonal elements $\sigma^2_{v_T}$ are the variances in the distance measurements (here supposed to be all equals) .

$$\mathbf{R}_{\mathrm{T}} = \begin{bmatrix} \sigma_{v_{\mathrm{T}}}^2 & 0 & 0 \\ 0 & \ddots & 0 \\ 0 & 0 & \sigma_{v_{\mathrm{T}}}^2 \end{bmatrix} \tag{18}$$

For the hyperbolic case, the covariance matrix of the noise is correlated [30].

$$\mathbf{R}_{\mathrm{T}} = \begin{bmatrix} \sigma_{v_{\mathrm{T}}}^2 & 0.5 \cdot \sigma_{v_{\mathrm{T}}}^2 & 0.5 \cdot \sigma_{v_{\mathrm{T}}}^2 \\ 0.5 \cdot \sigma_{v_{\mathrm{T}}}^2 & \ddots & 0.5 \cdot \sigma_{v_{\mathrm{T}}}^2 \\ 0.5 \cdot \sigma_{v_{\mathrm{T}}}^2 & 0.5 \cdot \sigma_{v_{\mathrm{T}}}^2 & \sigma_{v_{\mathrm{T}}}^2 \end{bmatrix} \tag{19}$$

Each sigma point is propagated using the odometry information in order to obtain the *a priori* sigma points estimation:

$$\boldsymbol{\chi}_{\mathrm{T},(n),k,j}^{-} = \begin{bmatrix} \chi_{\mathrm{T},(n),k,j,1} + \Delta d_{odo} \cdot \cos\bigl(\chi_{\mathrm{T},(n),k,j,3} + \Delta\theta_{odo}\bigr) \\ \chi_{\mathrm{T},(n),k,j,2} + \Delta d_{odo} \cdot \sin\bigl(\chi_{\mathrm{T},(n),k,j,3} + \Delta\theta_{odo}\bigr) \\ \chi_{\mathrm{T},(n),k,j,3} + \Delta\theta_{odo} \end{bmatrix}, j = 0, \dots, 2L \tag{20}$$

The values $\chi_{\mathrm{T},(n),k,j,1}$ and $\chi_{\mathrm{T},(n),k,j,2}$ are the components of the coordinates of the sigma point $j$ and $\chi_{\mathrm{T},(n),k,j,3}$ is the orientation.

The final step to finish with the prediction stage is to obtain the *a priori* estimation of the state vector and the *a priori* covariance matrix, using the transformed sigma points (20) according to the following expressions:

$$\widehat{\mathbf{X}}_{\mathrm{T},(n),k}^{-} = \sum_{j=0}^{2L} \mathbf{W}_{j}^{(d)} \cdot \boldsymbol{\chi}_{\mathrm{T},(n),k,j}^{-} \tag{21}$$

$$\mathbf{P}_{\widehat{\mathbf{X}}_{\mathrm{T},(n),k}}^{-} = \sum_{j=0}^{2L} \mathbf{W}_{j}^{(c)} \cdot [\boldsymbol{\chi}_{\mathrm{T},(n),k,j}^{-} - \widehat{\mathbf{X}}_{\mathrm{T},(n),k}^{-}] \cdot [\boldsymbol{\chi}_{\mathrm{T},(n),k,j}^{-} - \widehat{\mathbf{X}}_{\mathrm{T},(n),k}^{-}]^{T} \tag{22}$$

The values $\mathbf{W}_{j}^{(d)}$ and $\mathbf{W}_{j}^{(c)}$ are vectors whose values are obtained as follows:

$$\mathbf{W}_{0}^{(d)} = \frac{\lambda}{L + \lambda} \tag{23}$$

$$\mathbf{W}_{0}^{(c)} = \frac{\lambda}{L + \lambda} + 1 - \alpha^2 + \beta \tag{24}$$

$$\mathbf{W}_{j}^{(d)} = \mathbf{W}_{j}^{(c)} = \frac{1}{2(L + \lambda)}, j = 0, \dots, 2L \tag{25}$$

The constant $\alpha$ determines the dispersion of sigma points nearby the previous estimation of the global state vector its value is normally set to 0.001. The parameter $\beta$ is a value that represents the knowledge about the noise distribution (for Gaussian noise $\beta = 2$).

Regarding the update stage, we obtain the estimation of the observation set using the *a priori* sigma points (20). For the spherical case, the expression is:

$$\hat{\mathbf{Y}}_{\mathrm{T,o},k,j} = \begin{bmatrix} h_1(\boldsymbol{\chi}^-_{\mathrm{T},(n),k,j}) \\ \vdots \\ h_i(\boldsymbol{\chi}^-_{\mathrm{T},(n),k,j}) \\ \vdots \\ h_l(\boldsymbol{\chi}^-_{\mathrm{T},(n),k,j}) \end{bmatrix} = \begin{bmatrix} \hat{d}_{\hat{\rho}^-_{\mathrm{T},(n),k,j},\mathbf{B}_{\mathrm{T,o,1}}} \\ \vdots \\ \hat{d}_{\hat{\rho}^-_{\mathrm{T},(n),k,j},\mathbf{B}_{\mathrm{T,o},i}} \\ \vdots \\ \hat{d}_{\hat{\rho}^-_{\mathrm{T},(n),k,j},\mathbf{B}_{\mathrm{T,o},l}} \end{bmatrix} \tag{26}$$

The value $\hat{d}_{\hat{\rho}^-_{\mathrm{T},(n),k,j},\mathbf{B}_{\mathrm{T,o},i}}$ is the distance between the *a priori* position of the sigma point and the $i^{th}$ beacon of the $m^{th}$ GRULPS (T=G and o=m) or $n^{th}$ LRULPS (T=L and o=n).

The estimation of the observations in the hyperbolic case is:

$$\hat{\mathbf{Y}}_{\mathrm{T,o},k,j} = \begin{bmatrix} \hat{d}_{\hat{\rho}^-_{\mathrm{T},(n),k,j},\mathbf{B}_{\mathrm{T,o,2}}} - \hat{d}_{\hat{\rho}^-_{\mathrm{T},(n),k,j},\mathbf{B}_{\mathrm{T,o,1}}} \\ \vdots \\ \hat{d}_{\hat{\rho}^-_{\mathrm{T},(n),k,j},\mathbf{B}_{\mathrm{T,o},i}} - \hat{d}_{\hat{\rho}^-_{\mathrm{T},(n),k,j},\mathbf{B}_{\mathrm{T,o,1}}} \\ \vdots \\ \hat{d}_{\hat{\rho}^-_{\mathrm{T},(n),k,j},\mathbf{B}_{\mathrm{T,o},l}} - \hat{d}_{\hat{\rho}^-_{\mathrm{T},(n),k,j},\mathbf{B}_{\mathrm{T,o,1}}} \end{bmatrix} \tag{27}$$

At this point, it is possible to obtain the measurement estimations based on the transformed sigma points $\hat{\mathbf{Y}}_{\mathrm{T,o},k,j}$ and the weights vector defined previously.

$$\hat{\mathbf{Z}}_{\mathrm{T},(n),k} = \sum_{j=0}^{2L} \mathbf{W}_j^{(d)} \cdot \hat{\mathbf{Y}}_{\mathrm{T,o},k,j} \tag{28}$$

The covariance matrix related to the measurement estimations is:

$$\mathbf{P}_{\hat{\mathbf{Z}}_{\mathrm{T},(n),k}} = \sum_{j=0}^{2L} \mathbf{W}_j^{(c)} \cdot [\hat{\mathbf{Y}}_{\mathrm{T,o},k,j} - \hat{\mathbf{Z}}_{\mathrm{T},(n),k}] \cdot [\hat{\mathbf{Y}}_{\mathrm{T,o},k,j} - \hat{\mathbf{Z}}_{\mathrm{T},(n),k}]^T \tag{29}$$

The cross-correlation between the covariances of the *a priori* state estimation and the measurements is equal to:

$$\mathbf{P}_{\hat{\mathbf{X}}^-_{\mathrm{T},(n),k},\hat{\mathbf{Z}}_{\mathrm{T},(n),k}} = \sum_{j=0}^{2L} \mathbf{W}_j^{(c)} \cdot [\boldsymbol{\chi}^-_{\mathrm{T},(n),k,j} - \hat{\mathbf{X}}^-_{\mathrm{T},(n),k}] \cdot [\boldsymbol{\chi}^-_{\mathrm{T},(n),k,j} - \hat{\mathbf{X}}^-_{\mathrm{T},(n),k}]^T \tag{30}$$

At last, with the following expressions we can obtain the estimation of the global state vector at instant $k$ and the update of the covariance matrix:

$$\mathbf{K} = \mathbf{P}_{\hat{\mathbf{X}}_{T,(n),k}^-, \hat{\mathbf{Z}}_{T,(n),k}} \cdot \mathbf{P}_{\hat{\mathbf{X}}_{T,(n),k}^-, \hat{\mathbf{Z}}_{T,(n),k}}^{-1}$$
(31)

$$\hat{\mathbf{X}}_{T,(n),k} = \hat{\mathbf{X}}_{T,(n),k}^- + \mathbf{K} \cdot (\mathbf{Z}_{T,o,k} - \hat{\mathbf{Z}}_{T,o,k})$$
(32)

$$\mathbf{P}_{T,(n),k} = \mathbf{P}_{\hat{\mathbf{X}}_{T,(n),k}}^- - \mathbf{K} \cdot \mathbf{P}_{\hat{\mathbf{Z}}_{T,(n),k}} \cdot \mathbf{K}^T$$
(33)

### 3.4.2. H-Infinity Filter (H-∞)

This filter is similar to the EKF with a main difference in its minimization criterion. A brief explanation of the filter is shown below, since the expressions are developed in [29].

The prediction and the update stages particularized to the global or local state vectors estimation are as follows:

Prediction stage:

In this step, an "a priori" estimation of the state vector using the information of the previous iteration and the increments of distance and angle ($\Delta d_{odo,k}, \Delta \theta_{odo,k}$) is obtained:

$$\hat{\mathbf{X}}_{T,(n),k}^- = f(\hat{\mathbf{X}}_{T,(n),k-1}^-, \Delta d_{odo,k}, \Delta \theta_{odo,k})$$
(34)

Update stage:

The first step to obtain the final estimation of the state vector and its covariance matrix is the calculation of the matrix $\mathbf{S}_{T,(n),k}$ as:

$$\mathbf{S}_{T,(n),k} = (\mathbf{I} \cdot \gamma \cdot \mathbf{P}_{T,(n),k-1} + \mathbf{H}_{T,(n),k}^T \cdot \mathbf{R}_T^{-1} \cdot \mathbf{H}_{T,(n),k} \cdot \mathbf{P}_{T,(n),k-1})^{-1}$$
(35)

Where $\mathbf{I}$ is a third-order identity matrix, $\gamma$ is a predetermined attenuation noise level, $\mathbf{P}_{T,(n),k-1}$ is the covariance matrix of the state vector in the previous step, $\mathbf{H}_{T,(n),k}$ is the derivative of the distances or difference of distances between the current "a priori" position and the beacons of the GRULPS or the $n^{th}$ LRULPS with respect to the state vector; and $\mathbf{R}_T$ is the covariance matrix of the measured noise, being the same as the UKF filter defined in (18) and (19).

The following step is the derivation of the gain $\mathbf{K}$ using the expression (36).

$$\mathbf{K} = \mathbf{A}_{T,(n),k} \cdot \mathbf{P}_{T,(n),k-1} \cdot \mathbf{S}_{T,(n),k} \cdot \mathbf{H}_{T,(n),k}^{T} \cdot \mathbf{R}_{T}^{-1}$$

(36)

The matrix $\mathbf{A}_{T,(n),k}$ represents the derivative with respect to the state vector of the relationship of the states between the current and the previous iteration.

Finally, the state vector estimation and the covariance matrix are obtained by the expressions (37) and (38), where $\mathbf{Z}_{T,o,k}$ is the observation vector, $\hat{\mathbf{Z}}_{T,o,k}$ is the estimation of the observation vector using the "a priori" state vector, and $\mathbf{Q}_{T}$ is the process noise showed in (17).

$$\hat{\mathbf{X}}_{T,(n),k} = \hat{\mathbf{X}}_{T,(n),k}^{-} + \mathbf{K} \cdot (\mathbf{Z}_{T,o,k} - \hat{\mathbf{Z}}_{T,o,k})$$

(37)

$$\mathbf{P}_{T,(n),k} = \mathbf{A}_{T,(n),k} \cdot \mathbf{P}_{T,(n),k-1} \cdot \mathbf{S}_{T,(n),k} \cdot \mathbf{A}_{T,(n),k} \cdot \mathbf{Q}_{T}$$

(38)

### 3.5. Autocalibration based on a Transformation Vector

When the MR navigates inside the coverage areas of GRULPSs and LRULPSs, it is possible to obtain the transformation vector between a $n^{th}$ local reference system and the global reference system. Such transformation vector is obtained using several points of the global and local trajectories in the common coverage areas of two ULPSs. After obtaining the transformation vector related to the LRULPS $n$, we can estimate the $n^{th}$ local beacon coordinates $\hat{\mathbf{B}}_{L,n}^{l}$ referenced to the global reference system.

We have implemented two methods to obtain the transformation vector: a) using an analytical method from two points in the $n^{th}$ local trajectory and the corresponding two points in the global one; and b) using a numerical optimization from a set of points in each one of the reference (local and global) systems. The following subsections describe each method.

### 3.5.1. Analytical method

In this case, the transformation vector of the estimation of the position of the beacons in a LRULPS is obtained using the last two points of the global trajectory separated a configurable distance $d_{min}$ and the corresponding two points of the LRULPS whose beacons will be calibrated (Fig. 8).

The process of the calculation of the transformation vector is described as follows:

- The matrix $\mathbf{T_A}$ contains the two points $[x_{L,n,k}, y_{L,n,k}]$ and $[x_{L,n,k-t}, y_{L,n,k-t}]$ in the local trajectory separated at least $d_{min}$ distance. The points are included in the matrix following the form shows as:

$$\mathbf{T_A} = \begin{bmatrix} \hat{x}_{L,n,k} & -\hat{y}_{L,n,k} & 1 & 0 \\ \hat{y}_{L,n,k} & \hat{x}_{L,n,k} & 0 & 1 \\ \hat{x}_{L,n,k-t} & -\hat{y}_{L,n,k-t} & 1 & 0 \\ \hat{y}_{L,n,k-t} & \hat{x}_{L,n,k-t} & 0 & 1 \end{bmatrix}$$

(39)

- The two corresponding points in the global trajectory compose a vector defined as:

$$\mathbf{T_B} = \begin{bmatrix} \hat{x}_{G,n,k} \\ \hat{y}_{G,n,k} \\ \hat{x}_{G,n,k-t} \\ \hat{y}_{G,n,k-t} \end{bmatrix} \tag{40}$$

- Finally, the expression to obtain the transformation vector according to the LRULPS $n$ is equal to:

$$\mathbf{T_{C,n}} = \mathbf{T_A^{-1}} \cdot \mathbf{T_B} = \begin{bmatrix} T_{C,n,1} \\ T_{C,n,2} \\ T_{C,n,3} \\ T_{C,n,4} \end{bmatrix} \tag{41}$$

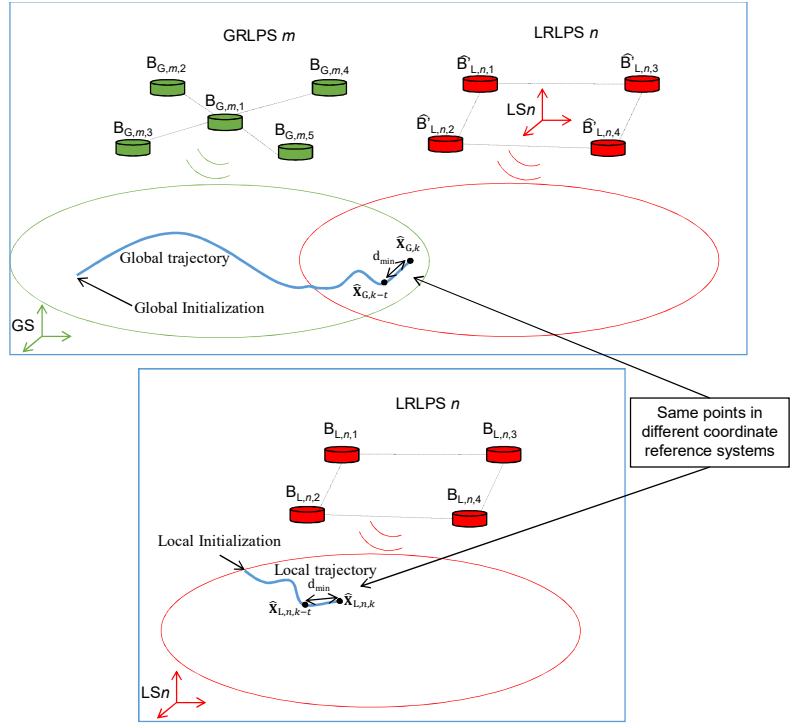

**Fig. 8.** Diagram of the LRULPS beacons' autocalibration based on an analytical method.

When the MR is in the last position of the common coverage area between both ULPSs, an average of the transformation vector parameters calculated along the time in which the MR was in the common area is obtained, $\overline{\mathbf{T}}_{\mathbf{C,n}}$. In this way, the robustness of the transformation process is increased.

Each beacon $i$ of the LRULPS $n$ is transformed using the components of the average transformation vector previously calculated:

$$\hat{x}'_{B,n,i} = x_{B,n,i} \cdot \overline{T}_{C,n,1} - y_{B,n,i} \cdot \overline{T}_{C,n,2} + \overline{T}_{C,n,3} \tag{42}$$

$$\hat{y}'_{B,n,i} = y_{B,n,i} \cdot \overline{T}_{C,n,1} + x_{B,n,i} \cdot \overline{T}_{C,n,2} + \overline{T}_{C,n,4} \tag{43}$$

The value $\overline{T}_{C,n,j}$ is the $j^{th}$ component of the average transformation vector.

### 3.5.2. Numerical method

In this method, we use a set of more than two points per reference system (Fig. 9) in order to obtain the transformation vector between coordinate systems.

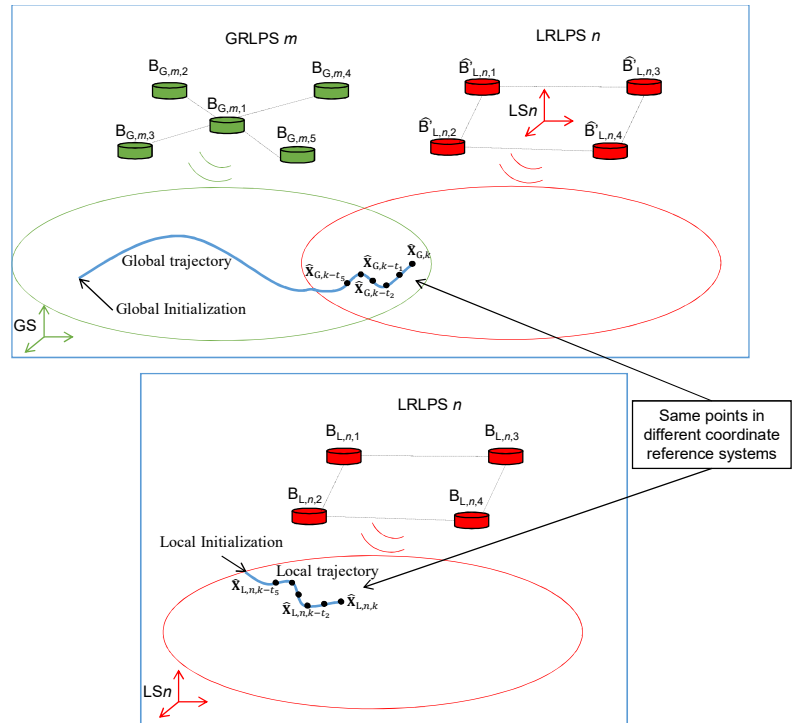

**Fig. 9.** Diagram of the LRULPS beacons' autocalibration based on an analytical method.

This method consists of determining the transformation vector that minimizes the mean distance error between a number of points in the global trajectory and the equivalent transformed points obtained from the LRULPS using this transformation vector. The set of points is selected taking into account that any pair of them has a separation greater than a pre-established value, $d_{min}$.

The numerical minimization has been performed using the Matlab function "fminunc".. The number of points selected to estimate the transformation vector is also a configurable parameter.

If the transformation vector is $\mathbf{T}_{C,n}$, with four components as in (41), all the points in the local reference system can be transformed to the global one. For instance, the point $(\hat{x}_{L,n,k}, \hat{y}_{L,n,k})$ in the local reference system, can be transformed to the global reference system as $(\hat{x}'_{G,n,k}, \hat{y}'_{G,n,k})$ using (44) and (45):

$$\hat{x}'_{G,n,k} = \hat{x}_{L,n,k} \cdot T_{C,n,1} - \hat{y}_{L,n,k} \cdot T_{C,n,2} + T_{C,n,3} \tag{44}$$

$$\hat{y}'_{G,n,k} = \hat{y}_{L,n,k} \cdot T_{C,n,1} + \hat{x}_{L,n,k} \cdot T_{C,n,2} + T_{C,n,4} \tag{45}$$

The mean error (ME) between all the corresponding points in the global coordinate system (the points obtained with the global filter and the ones obtained with the transformation from de local reference) is:

$$ME = \frac{1}{T} \cdot \sum_{t=0}^{T-1} \left( \sqrt{\left(\hat{x}'_{G,n,k-t} - \hat{x}_{G,k-t}\right)^2 + \left(\hat{y}'_{G,n,k-t} - \hat{y}_{G,k-t}\right)^2} \right) \tag{46}$$

Where $T$ is the number of the selected points to obtain the transformation vector.

And the values of de components of $\mathbf{T}_{C,n}$ are the ones that minimize the ME:

$$\mathbf{T}_{C,n} = \underset{[T_{c,n,j}], \ j=1,2,3,4}{\mathbf{argmin}} (ME) \tag{47}$$

Finally, as in the previous case, the position of the beacons of the LRULPS $n$, referenced to the global system, are obtained using the expressions (42) and (43).

### 3.6. Use of the Inverse Trajectory

After navigating through the coverage area of several LRULPS, when the MR arrives to a GRULPS the estimation error of the trajectory is drastically reduced since this GRULPS has been previously calibrated. A way to improve the estimation of the LRULPS beacons is using a new virtual trajectory of the MR starting from the final position (inside the GRULPS coverage area) and going backwards running the algorithm again if all the measurements were saved. In the following, we denote that as "inverse trajectory" and its obvious effect is an improvement of the last LRULPS calibrated as they are the first in this virtual inverse trajectory.

## 4. Simulated Results

We have simulated a trajectory intended for selfcalibrating the beacons of several ($n$=7) LRULPSs.
Fig. 10 shows the representation of this trajectory as well as the distribution of GRULPSs and LRULPSs (note that they are projected on the floor).

The value of the noise related to the odometer (increments of distance and angle between successive calculations) and US signals (distances measured from beacons to receiver) are set the following typical deviations: $\sigma_{\Delta d_{odo}} = 0.03m$, $\sigma_{\Delta \theta_{odo}} = 0.02m$ and $\sigma_v = 0.005m$.
These values are approximate to a real noise conditions and obtained heuristically.

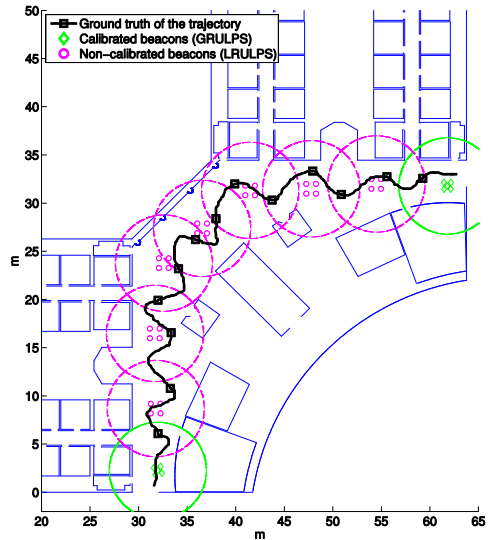

**Fig. 10.** Simulated trajectory to autocalibrate the beacons.

*4.1. Analytical Method*

As stated before, this method uses two points of the $n^{th}$ local trajectory and the corresponding two points of the global path. The parameter $d_{min}$ represents the distance between the two points selected. A low value of this distance is not recommendable since the error in the estimation of the transformation vector in that case could be high due to bad geometric configurations. According to preliminary simulations we have set this value to 2.5m. Fig. 11 shows a CDF of the error in the autocalibration of all the beacons shown in Fig. 10 for different variations of this parameter ($d_{min}$). As it can be seen, low ($0.5m$ *or* $1,5m$) and high ($5m$) values of this parameter present a worse performance related to the error in the calibration of the beacons than a middle value ($2.5m$), in which the error is lower than 0.4m in the 95% of cases.

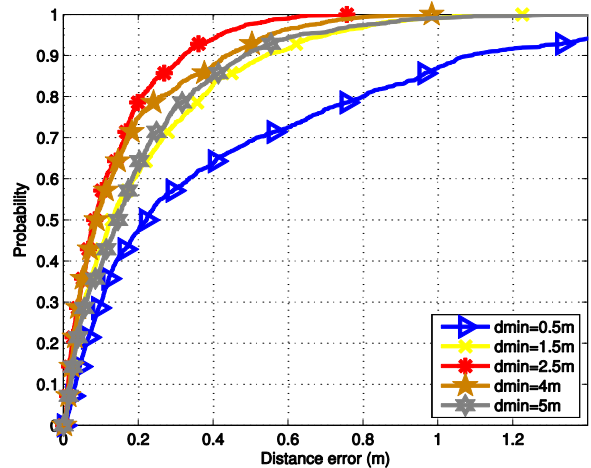

**Fig. 11.** CDF of the distance error related to all estimated beacons based on several values of $d_{min}$.

Fig. 12 shows the simulated results for the spherical (left image) and hyperbolic (right image) cases including the correction of the three last LRULPSs using the inverse trajectory.

As it can be observed there is an incremental error in the estimated trajectories based on the filters EKF, UKF and H-∞ when the MR is moving between LRULPSs. This error is almost reset when the MR is inside another GRULPS coverage area and the estimation of the trajectory in that case is quite close to the ground truth.

Regarding the autocalibration of the LRULPS (Fig. 13), the error is higher in the LRULPS 5 that is placed around the middle of the trajectory between the two GRULPSs considered. For the spherical case the mean error is lower than 0.2m and for the hyperbolic case less than 0.4m.

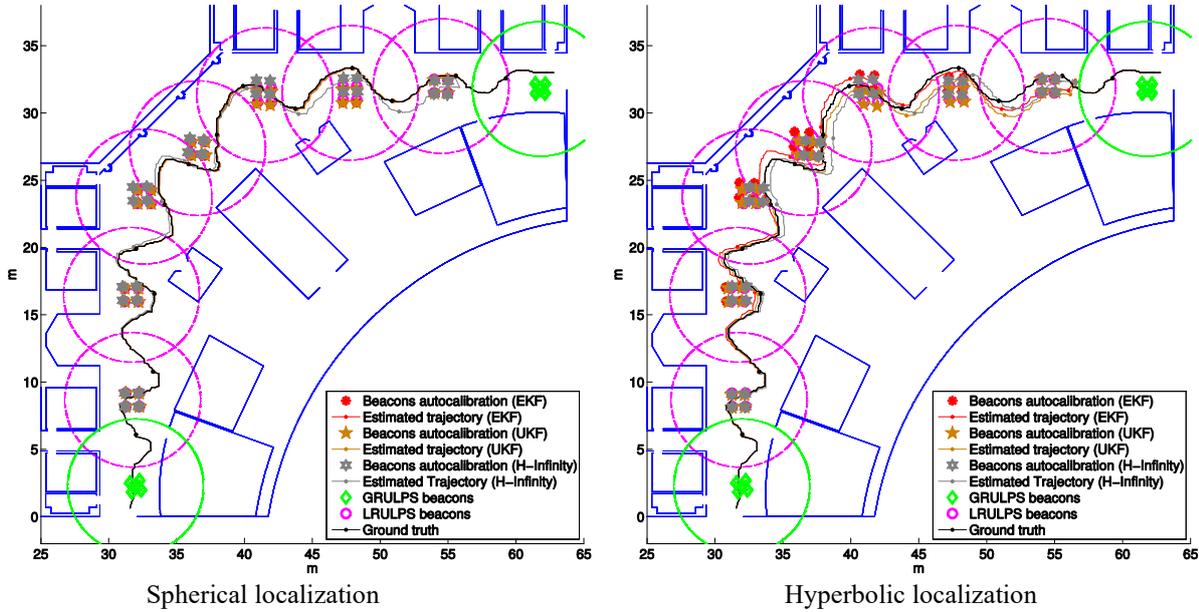

Spherical localization        Hyperbolic localization

**Fig. 12.** Single simulation of SCAN based on the Analytical Method.

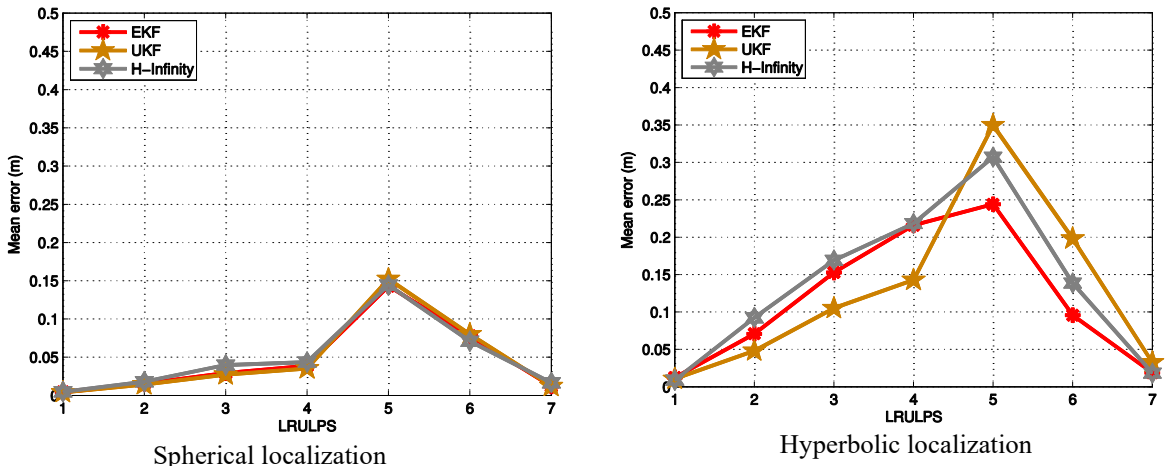

Spherical localization        Hyperbolic localization

**Fig. 13.** Autocalibration average errors after 100 iterations.

## 4.2. Numerical Method

As described before, this method considers more than two points for each reference system. The transformation vector is obtained minimizing a cost function (46).

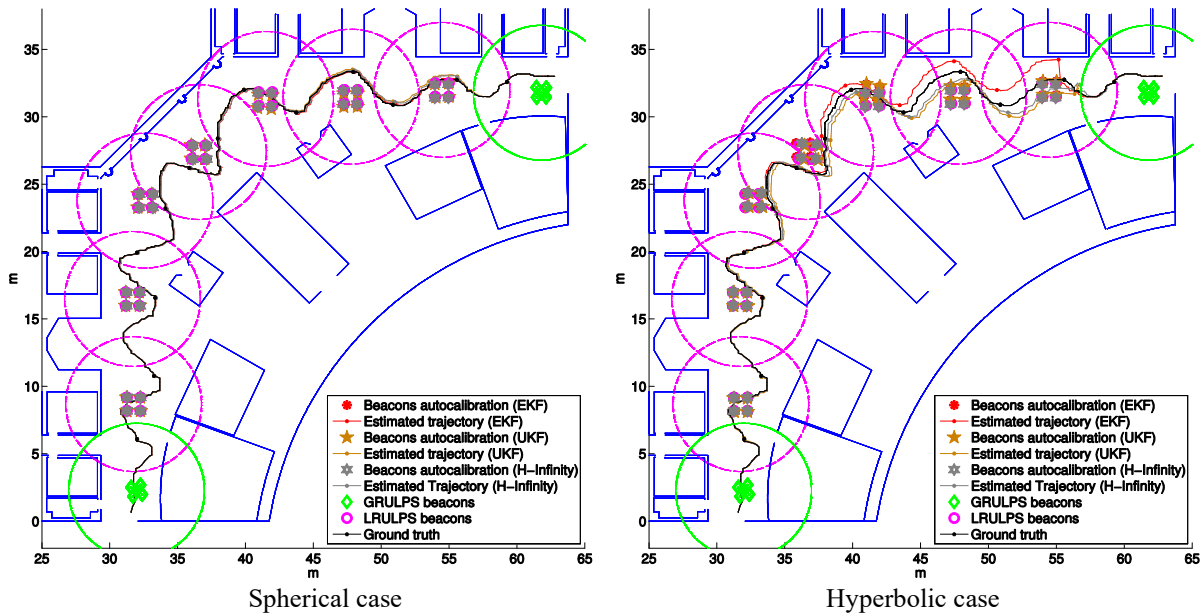

Spherical case                    Hyperbolic case

**Fig. 14.** Single simulation of SCAN based on the Numerical Method.

The advantage of the use of more than two points per reference system is the reduction of the variance while the positions of the beacons are being calculated while the MR navigates. The results are similar to the Analytic Method achieving errors lower than 0.15m and 0.35m for the spherical and hyperbolic cases as Fig. 14 and Fig. 15 show.

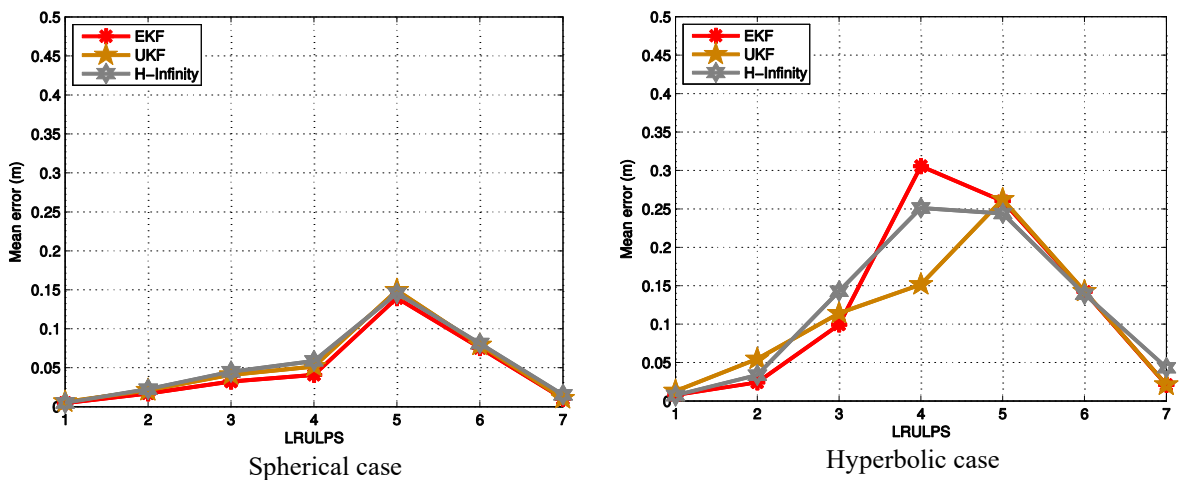

Spherical case                    Hyperbolic case

**Fig. 15.** Autocalibration average errors after 100 iterations.

## 5. Experimental Results

In order to verify the simulations, we have performed a real test in a part of the same environment than the one used in simulations (Fig. 16), although in this case with 3 LRULPSs and 2 GRULPSs.

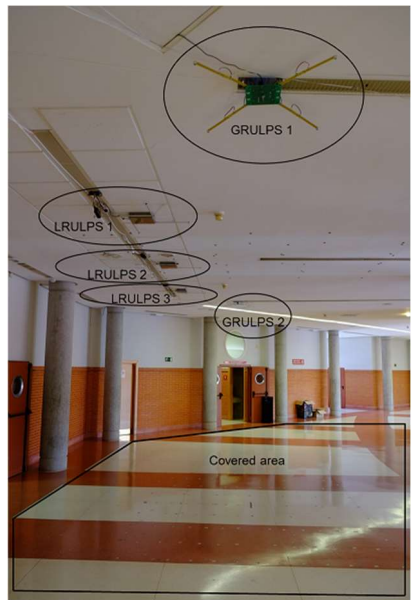

**Fig. 16.** Real tests scenario and ULPSs location.

The localization area is covered by five ULPSs whose transducers are ProWave 328ST160 [31]. Fig. 17 shows a detail of the structure and implementation of the ULPS.

All transducers emit a Kasami code of 255 bit with a BPSK modulation and a carrier frequency of 41.66KHz. The control unit is based on a FPGA Xilinx Zynq-7000 [32].

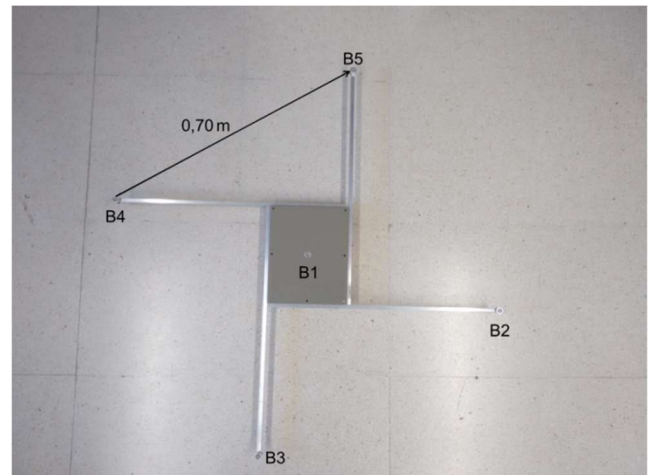

**Fig. 17.** ULPS structure.

The receiver is composed of a microphone, type Brüel & Kjaer 4939, and a digitizer, Avisoft-Bioacustics CM16/CMPA, connected to a laptop arranged on a MR Pioneer-3DX [33] (see Fig. 18).

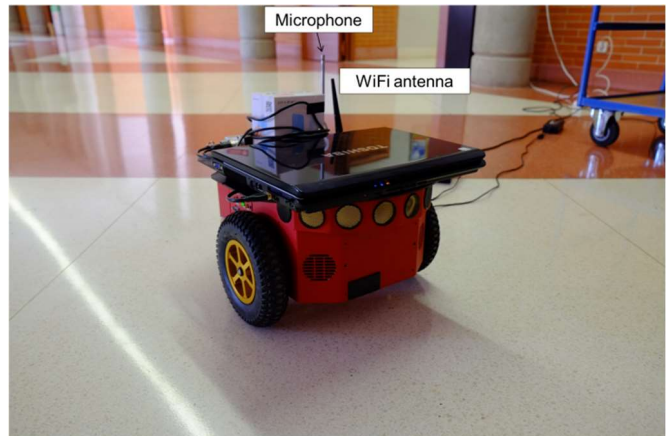

**Fig. 18.** MR Pioneer 3DX.

Fig. 19 shows the approximate real trajectory and coverages of each ULPS (different colors) obtained with a Gauss-Newton algorithm for the hyperbolic case. As it can be observed in the Figure the common coverage between consecutive ULPSs is around 1 meter, so in this case the parameter $d_{min}$ was set to 0.5m.

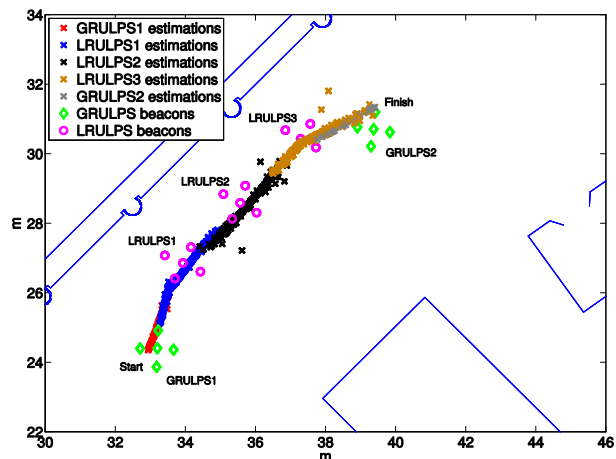

**Fig. 19.** Position estimation based on Gauss-Newton algorithm for each ULPS.

Results regarding to the analytical and numerical methods for the SCAN process are shown in the following subsections.

## 5.1. Analytical Method Results

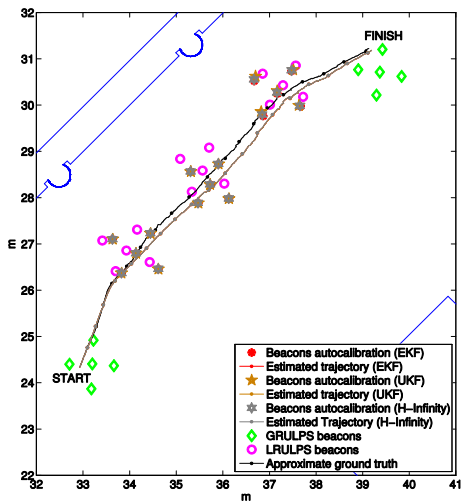
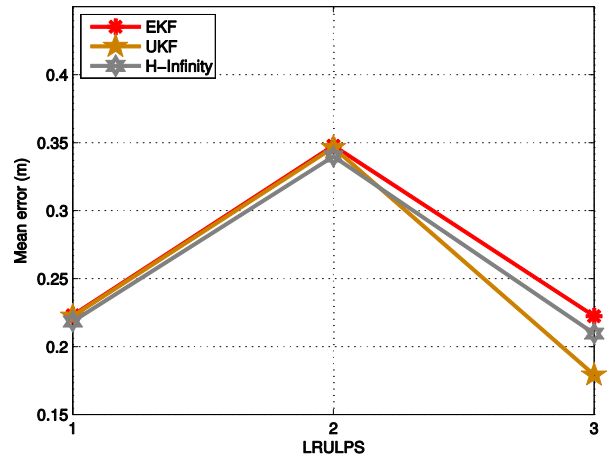

SCAN visual results                          Autocalibration average errors

**Fig. 20.** Real SCAN results based on the Analytical Method (hyperbolic positioning).

Fig. 20 shows the results of the SCAN proposal based on the Analytical Method. As is showed in simulations, the worst beacons' estimation is in the LRULPS placed on the middle between the GRULPSs with an average error lower than 0.33m for the H-∞ filter and around 0.35m for the EKF and UKF filters.

## 5.2. Numerical Method Results

Fig. 21 shows the SCAN results based on the Numerical Method.

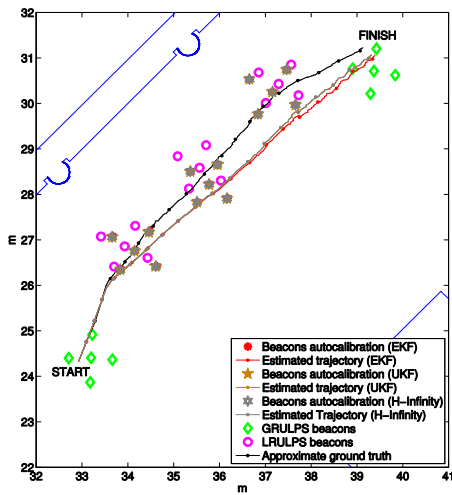
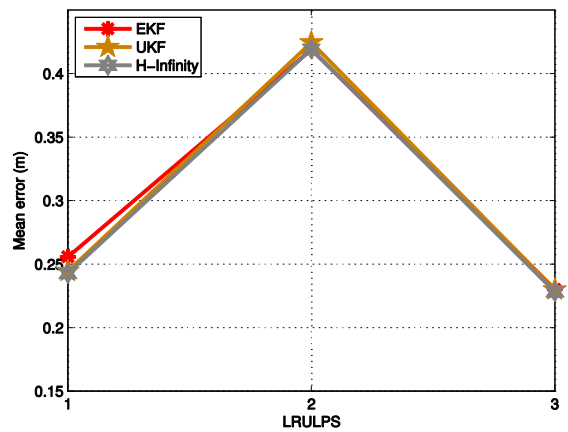

SCAN visual results                          Autocalibration average errors

**Fig. 21.** Real SCAN results based on the Numerical Method.

The maximum error is around 0.45m for all filters in the second LRULPS. The results are worse than the Analytical Method since the algorithm uses the same two points than the Analytical case and other extra points for each reference system in order to estimate the transformation vector and it is possible that these other points contain higher errors than the main two points.

## 6. Conclusions

This paper has described a Simultaneous Calibration and Navigation (SCAN) process used by a MR to navigate with the help of multiple Ultrasonic Local Positioning Systems whose beacons' position are also estimated. We have considered some Globally Referenced Ultrasonic Local Positioning Systems (GRULPS) whose beacons are previously calibrated (the position of the beacons are known with respect to the global map) and placed in the entrances of the environment. Additionally, a bigger set of Locally Referenced Ultrasonic Local Positioning Systems (LRULPSs), whose beacons are uncalibrated, cover the rest of the navigation area. The proposal permits to navigate and autocalibrate the beacons simultenously using a MR with odometry.

The algorithm is based on the use of a filter for each reference system (one for each one of the LRULPSs) to estimate the trajectory of the MR inside the local reference area of each ULPS. The estimation of the position of the beacons of the LRULPSs is carried out obtaining a transformation vector that converts the coordinates of the local reference system to the global one. This transformation vector is obtained using two points of the local trajectory and the global one if the resolution method is analytic; or more than two points in case of using a numerical method. Both methods achieve mean errors in the estimation of the beacons' position lower than 0.2m and 0.5m for the spherical and hyperbolic cases, respectively, using simulations. Real experiments (for the hyperbolic case) give a mean error of 0.45m for the worst LRULPS estimation. The analytic method is simpler than the numerical one since it is required less positions and an easier computation to obtain the transformation vector.

Future research lines for improving the accuracy and robustness of these autocalibration systems can be: the use of more than one robot achieving a beacons' calibration based on their collaboration, exporting this system to portable devices for its orientation to people navigation, in which the uncalibrated beacons can be estimated by a high number of users with their localization applications in the mobile phones, or adding more information to the algorithm such as a map of the environment.

## Acknowledgments

This work has been possible thanks to the Spanish Ministry of Economy and Competitiveness (TARSIUS project, ref. TIN2015-71564-C4-1-R).